\documentclass[12pt,preprint]{aastex}
\usepackage[pdftex]{epsfig}
\usepackage{color}

\newcommand{\Grad}{$^{\circ}$}

\shortauthors{T.~Antoni et al. (KASCADE Collaboration) }
\shorttitle{Large scale cosmic-ray anisotropy with KASCADE}

\begin{document}

\title{Large scale cosmic-ray anisotropy with KASCADE}

\author{T.Antoni\altaffilmark{1},
        W.\,D.~Apel\altaffilmark{2},
        A.F.~Badea\altaffilmark{2,6},
        K.~Bekk\altaffilmark{2},
        A.~Bercuci\altaffilmark{2,6},
        H.~Bl\"umer\altaffilmark{2,1},
        H.~Bozdog\altaffilmark{2},
        I.\,M.~Brancus\altaffilmark{3},
        C.~B\"uttner\altaffilmark{1},
        K.~Daumiller\altaffilmark{1},
        P.~Doll\altaffilmark{2},
        R.~Engel\altaffilmark{2},
        J.~Engler\altaffilmark{2},
        F.~Fe{\ss}ler\altaffilmark{2},
        H.\,J.~Gils\altaffilmark{2},
        R.~Glasstetter\altaffilmark{1,7},
        A.~Haungs\altaffilmark{2},
        D.~Heck\altaffilmark{2},
        J.\,R.~H\"orandel\altaffilmark{1},
        K.-H.~Kampert\altaffilmark{1,2,7},
        H.\,O.~Klages\altaffilmark{2},
        G.\,Maier\altaffilmark{2,9},
        H.\,J.~Mathes\altaffilmark{2},
        H.\,J.~Mayer\altaffilmark{2},
        J.~Milke\altaffilmark{2},
        M.~M\"uller\altaffilmark{2},
        R.~Obenland\altaffilmark{2},
        J.~Oehlschl\"ager\altaffilmark{2},
        S.~Ostapchenko\altaffilmark{1,8},
        M.~Petcu\altaffilmark{3},
        H.~Rebel\altaffilmark{2},
        A.~Risse\altaffilmark{5},
        M.~Risse\altaffilmark{2},
        M.~Roth\altaffilmark{1},
        G.~Schatz\altaffilmark{2},
        H.~Schieler\altaffilmark{2},
        J.~Scholz\altaffilmark{2},
        T.~Thouw\altaffilmark{2},
        H.~Ulrich\altaffilmark{2},
        J.~van Buren\altaffilmark{2},
        A.~Vardanyan\altaffilmark{4},
        A.~Weindl\altaffilmark{2},
        J.~Wochele\altaffilmark{2},
        and J.~Zabierowski\altaffilmark{5} \\ (The KASCADE Collaboration)
       }

\altaffiltext{1}{Institut f\"ur Experimentelle Kernphysik, Universit\"at
        Karlsruhe, 76021~Karlsruhe, Germany}
\altaffiltext{2}{Institut f\"ur Kernphysik, Forschungszentrum Karlsruhe,
             76021~Karlsruhe, Germany}
\altaffiltext{3}{National Institute of Physics and Nuclear Engineering,
             7690~Bucharest, Romania}
\altaffiltext{4}{Cosmic Ray Division, Yerevan Physics Institute,
             Yerevan~36, Armenia}
\altaffiltext{5}{Soltan Institute for Nuclear Studies,
             90950~Lodz, Poland}
\altaffiltext{6}{on leave of absence from NIPNE, Bucharest}
\altaffiltext{7}{now at: Universit\"at Wuppertal, 42119 Wuppertal, Germany}
\altaffiltext{8}{on leave of absence from Moscow State University, Moscow, Russia}
\altaffiltext{9}{corresponding author, email: gernot.maier@ik.fzk.de}


\begin{abstract}
The results of an analysis of the large scale anisotropy 
of cosmic rays in the PeV range are presented.
The Rayleigh formalism is applied to the right ascension 
distribution of extensive air showers measured by the KASCADE
experiment.
The data set contains about $10^8$ extensive air showers
in the energy range from 0.7 to 6~PeV.
No hints for anisotropy are visible in the right ascension
distributions in this energy range.
This accounts for all showers as well as for subsets containing showers 
induced by predominantly light respectively heavy primary particles.
Upper flux limits for Rayleigh amplitudes are determined to be 
between $10^{-3}$ at 0.7~PeV and $10^{-2}$ at 6~PeV primary energy.
\end{abstract}


\section{Introduction}

The arrival direction of charged cosmic rays with primary energies
between several hundred TeV and 10~PeV is remarkably isotropic.
A possible anisotropy  would reflect the general pattern of
propagation of cosmic rays in the galactic environment.
Model calculations,  e.g.~of \cite{Candia:2003}
show that diffusion of cosmic rays
in the galactic magnetic field can result in an anisotropy on a scale of
$10^{-4}$ to $10^{-2}$ depending on particle energy and
strength and structure of the galactic magnetic field.
The diffusion is rigidity dependent, the cited model calculation
reports roughly a factor of five to ten larger anisotropy for
protons than for iron primary particles with the same energy.
This rigidity-dependent diffusion is one of several explanations of the
steepening in the cosmic ray energy spectrum at around
4~PeV.
Another class of models explains this so-called knee in the energy
spectrum 
as a result of a change in the acceleration efficiency of the
source (e.g.~\cite{Lagage:1983}).
There is no change in anisotropy at the knee expected from
these models, while the models based on diffusion should
result in an increase at about~4 PeV.
Anisotropy measurements give, in addition to the 
measurements of mass dependent energy spectra, valuable 
information for the discrimination between models
explaining the knee in the cosmic-ray energy spectrum.

Due to the small anisotropy expected a
large data sample is necessary.
The flux of cosmic rays in the PeV energy range is too low
for direct measurements by experiments on satellites or balloons.
Ground based experiments with large collecting areas measuring
the secondary products of the interaction of the
primary cosmic rays with Earth's atmosphere are presently the only
way to collect a suitable amount of events.
Few statistically significant anisotropies were reported
from extensive air shower experiments in the last two decades. 
EAS-TOP \citep{EASTOP96}
published an amplitude of $(3.7\pm 0.6)\times 10^{-4}$ at
$E_0\approx 200$~TeV.
The Akeno experiment \citep{Kifune:1986}
reported results of about $2\times 10^{-3}$ at about 5 to 10~PeV.
An overview of experimental results can be found in 
\citep{Clay:1997}.

In the following, the large scale cosmic-ray anisotropy is studied by application
of the Rayleigh formalism to data of the KASCADE air shower experiment.
The two-dimensional distribution of arrival directions of cosmic rays is
reduced to one coordinate due to the limited field of view and the small
amplitudes expected from theory and previous observations.
A first order approximation of the multipole expansion of the arrival
directions of cosmic rays
is a harmonic analysis of the right ascension values of extensive
air showers.
The Rayleigh formalism gives the amplitude $A$
and phase $\Phi$ of the first harmonic,
and additionally the probability $P$ for detecting a spurious amplitude due to 
fluctuations from a sample of $n$ events which are
drawn from a uniform distribution \citep{Mardia:1999}:
\begin{eqnarray}
&A& = \sqrt{C^2+S^2}, \quad \Phi = \arctan \frac{S}{C} \label{equ:Ray} \\
&S& = \frac{2}{n}\sum_{i=1}^{n} \sin \alpha_i, \  C = \frac{2}{n}\sum_{i=1}^{n} \cos \alpha_i \\
&P&(>A)=\exp{(-nA^2/4)}+O(n^{-2}) \label{equ:RayP} 
\end{eqnarray}
The sum includes $n$ right ascension values $\alpha_i$.
Studies of higher harmonics are very limited as the expected
amplitudes are too small compared to the statistical fluctuations 
of the data sets available.


In this article an analysis of data from the KASCADE experiment is presented,
which is described in the following section.
The data selection procedures, including an enrichment of light and heavy primary
particles are presented  in sections~3 and~4.
Section~5 describes the corrections applied to the shower rates depending on
atmospheric ground pressure and
temperature.
The main results, i.e.~the Rayleigh amplitudes for all showers as well
as for the mass enriched samples can be found in section~6.

\section{KASCADE - experimental setup and data reconstruction}

The extensive air shower experiment
KASCADE (KA{\em rlsruhe} S{\em hower} C{\em ore and} A{\em rray}
DE{\em tector}) is located at Forsch\-ungs\-zentrum Karlsruhe,
Germany ($8.4^{\mathrm{o}}$~E, $49.1^{\mathrm{o}}$~N) 
at \mbox{110~m} a.s.l.
corresponding to an average vertical atmospheric depth of \mbox{1022 g/cm$^2$}.
KASCADE measures the electromagnetic, muonic, and hadronic
components of air showers with three major detector systems: a large field
array, a muon tracking detector, and a central detector \citep{KASCADE03}.

In the present analysis data from the 200$\times$200~m$^2$ scintillation
detector array are used.
The 252 detector stations are uniformly spaced on a square grid of 13~m.
The stations are organized in 4$\times$4 electronically
independent clusters with 16 stations in the 12 outer and 15 stations in the
four inner clusters.
The stations in the inner/outer clusters contain four/two
liquid scintillator detectors covering a total area of 490~m$^2$.
Additionally, plastic scintillators are mounted below an absorber
of 10~cm of lead and 4~cm of iron in the 192 stations of the outer clusters
(622~m$^2$ total area).
The absorber corresponds to 20 electromagnetic radiations lengths
entailing a threshold for vertical muons of 230~MeV.
This configuration allows the measurement of the electromagnetic
and muonic components of extensive air showers.
The number of electrons ($N_e$) and muons ($N_{\mu,tr}$) in a shower, 
the position of the shower core and the shower direction
are determined in an iterative shower reconstruction procedure.
The 'truncated' muon number ($N_{\mu,tr}$) denotes the number 
of muons in the distance range from
40 to 200~m to the shower core.
Shower directions are determined without assuming a fixed
geometrical shape of the shower front by evaluating the arrival times of
the first particle in each detector and the total particle number
per station.
The angular resolution for zenith angles less than 40\Grad is
0.55\Grad \ for small showers and 0.1\Grad \ for showers with
electron numbers of $\log_{10} N_e \geq 6$.

The detector array reaches full detection efficiency
for extensive air showers with electron numbers $\log_{10} N_e > 4$
corresponding to a primary energy of about $(6-9)\times 10^{14}$ eV.
This is defined by a detector multiplicity condition resulting
in a trigger rate of about 3~Hz.
The data set for the following analysis contains 
$10^8$ events recorded in 1600 days between
May 1998 and October 2002.


\section{Data selection}
\label{sec:cuts}

Because of the very small amplitudes expected a very careful data selection is necessary.
Contributions from amplitudes in local solar time can cause spurious signals
in sidereal time.
This leakage is due to the very small difference in daylength
between a solar and a sidereal day ($\Delta t=236$ s).
Amplitudes in solar time can be caused by
variation of atmospheric ground pressure and temperature,
and will be corrected for (Chapter 5).
To minimize these spurious effects, several
cuts are applied to the measured showers in order to
enhance data quality.
In the following, rates are determined in time intervals
of half an hour (in sidereal time $\approx 1795$ SI seconds).
In detail the selection criteria are: 
\begin{enumerate}
\item To ensure  reconstruction quality, only showers well inside
the detector field with a maximum distance to its center 
of 91~m and with zenith angles smaller than $40$\Grad \ have been used.
The latter cut restricts the visible sky 
to the declination band $9$\Grad $<\delta<89$\Grad .
\item More than 249 out of 252 detector stations have to be in 
working condition.
\item Sudden changes of the rate are detected by testing the uniformity of
the rate as a function of time for each sidereal day.
No deviations of the rates from the mean rate larger than 4$\sigma$
determined over the whole measurement time are allowed.
\item Only sidereal days with continuous data taking are used.
\item The array has to be fully efficient (100\%) for EAS detection.
Simulations show that this is the case for electron numbers $\log_{10} N_e > 4$.
\end{enumerate}

After application of these quality cuts about 20\% of the
showers from the initial data set are left.
In total 269 out of 1622 sidereal days with continuous data taking are used
in the following analysis.
The seasonal distribution of these days is as follows:
114 days in spring (February-April), 18 in summer (May-July),
77 in autumn (August-October), and 60 in winter (November-January).

\section{Enrichment of light and heavy primaries}

\begin{figure}
\plottwo{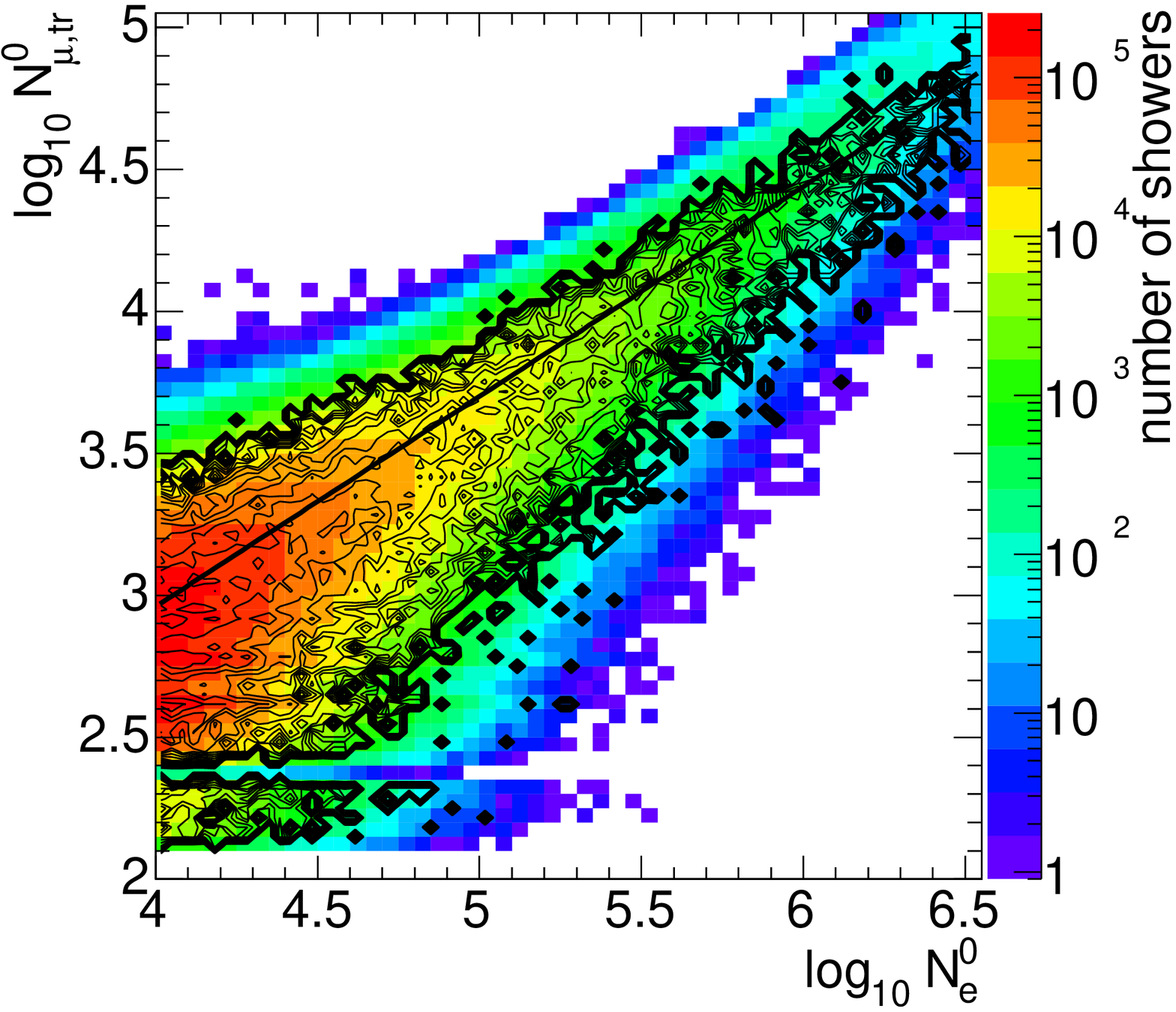}{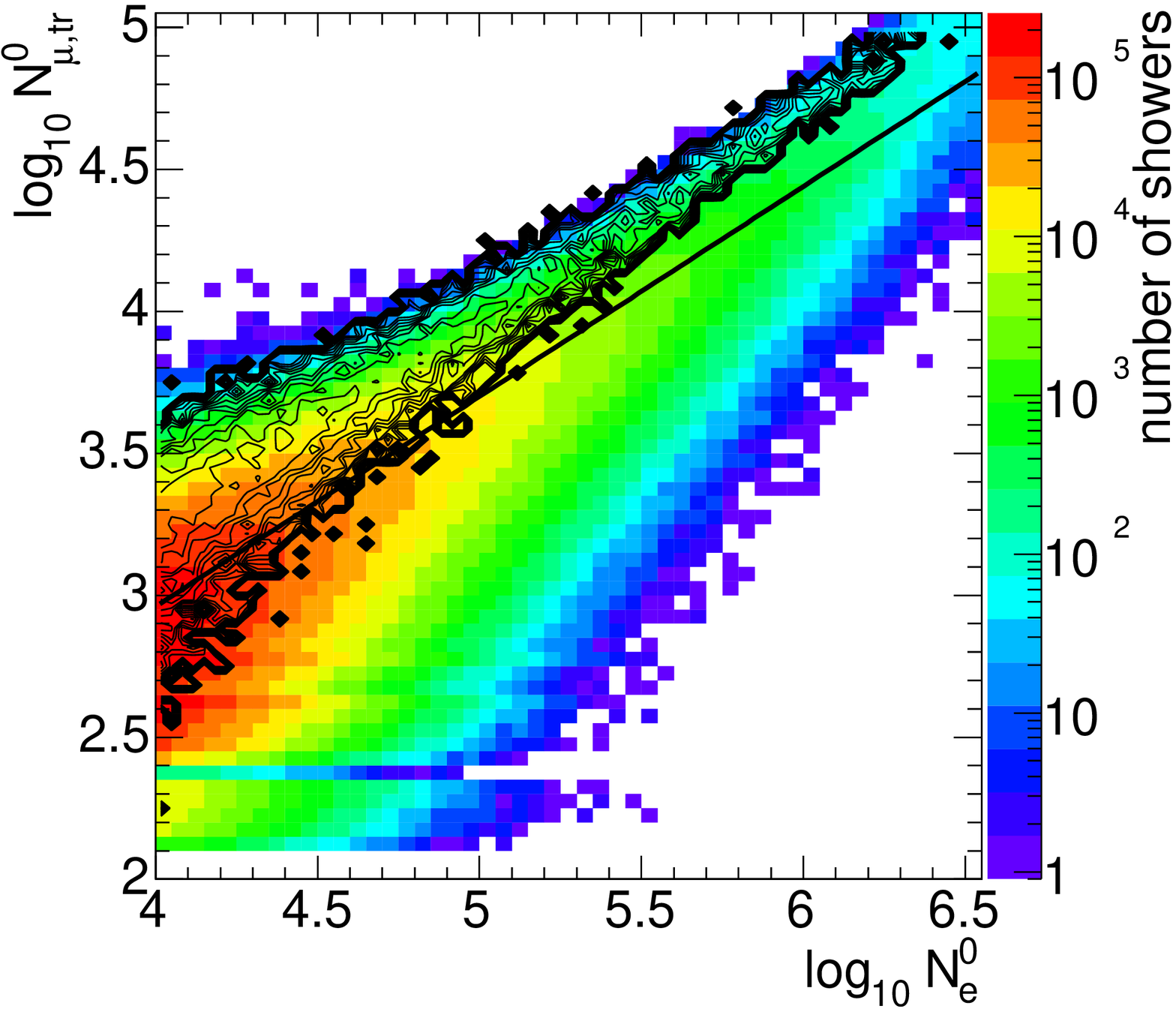}
\caption{\label{fig:gammaSep}
{Number of muons versus number of electrons ($\log_{10} N_{\mu,tr}^0$ vs. $\log_{10} N_e^0$)
of showers measured with KASCADE (shaded area).
Simulated air showers induced by primary protons (left-hand side) or iron nuclei (right-hand side)
are superimposed (contour lines).
The straight lines in both figures indicate the separation between light and heavy primaries according to equation \ref{equ:gammasSepCut}.
}}
\end{figure}

To evaluate the dependence of a possible Rayleigh amplitude on primary
energy and mass, the data set is divided by a simple cut in
the $\log_{10} N_{\mu,tr}$-$\log_{10} N_e$ plane into two sets.
Simulation studies show,  that showers
initiated by light primary particles are predominately electron rich
while those from heavy primaries are electron poor \citep{Antoni:2002}.
The extensive air showers are simulated utilising the CORSIKA package \citep{Heck:1998}.
The QGSJet-model \citep{Kalmykov:1997} is used for hadronic interactions
above $E_{Lab}>80$~GeV, GHEISHA \citep{Fesefeldt:1985} for interactions below
this energy.
The electromagnetic cascades are simulated by EGS4 \citep{Nelson:1985}.
The shower simulation is followed by a 
detector simulation based on GEANT \citep{geant}.
A power law with constant spectral index of $\gamma=-2$
is used in the simulations.
Figure \ref{fig:gammaSep} shows the distribution of
muon number versus electron number ($\log_{10} N_{\mu,tr}^0$-$\log_{10} N_e^0$)
of showers measured with KASCADE and from the mentioned simulations of proton- and iron-induced showers.
There are several reasons for the differences between the measured and the simulated distributions.
The chemical composition of cosmic rays consists of more than two components, the slope of the
energy spectrum of the cosmic rays is different in the simulations and the number of measured showers
exceeds by far the number of simulated showers.

A separation between light and heavy primaries can be
expressed by the following ratio:
\begin{equation}
\label{equ:gammasSepCut}
\log_{10} N_{\mu,tr}^0 / \log_{10} N_e^0 = 0.74
\end{equation}
The electron number $ N_e$ and the truncated muon number $N_{\mu,tr}$
are zenith angle corrected to $\Theta=0$\Grad \ using the attenuation law:
\begin{eqnarray}
N_e^0 &=& N_e \cdot \exp{ \left(X_0/\Lambda_e(\mathrm{sec}\Theta-1)\right)} \\ 
N_{\mu,tr}^0 &=& N_{\mu,tr} \cdot \exp{\left(X_0/\Lambda_{\mu,tr}(\mathrm{sec}\Theta-1)\right)} 
\end{eqnarray}
with the attenuation lengths $\Lambda_{N_e}=175$~g/cm$^2$ and
$ \Lambda_{N_{\mu,tr}}=823$~g/cm$^2$ \citep{Antoni:2003}.

This separation neglects the large fluctuations,
especially of proton initiated showers.
It also neglects the very different relative abundances of light to heavy primaries
in cosmic rays.

\section{Correction for atmospheric ground pressure and temperature}

\begin{figure}
\plottwo{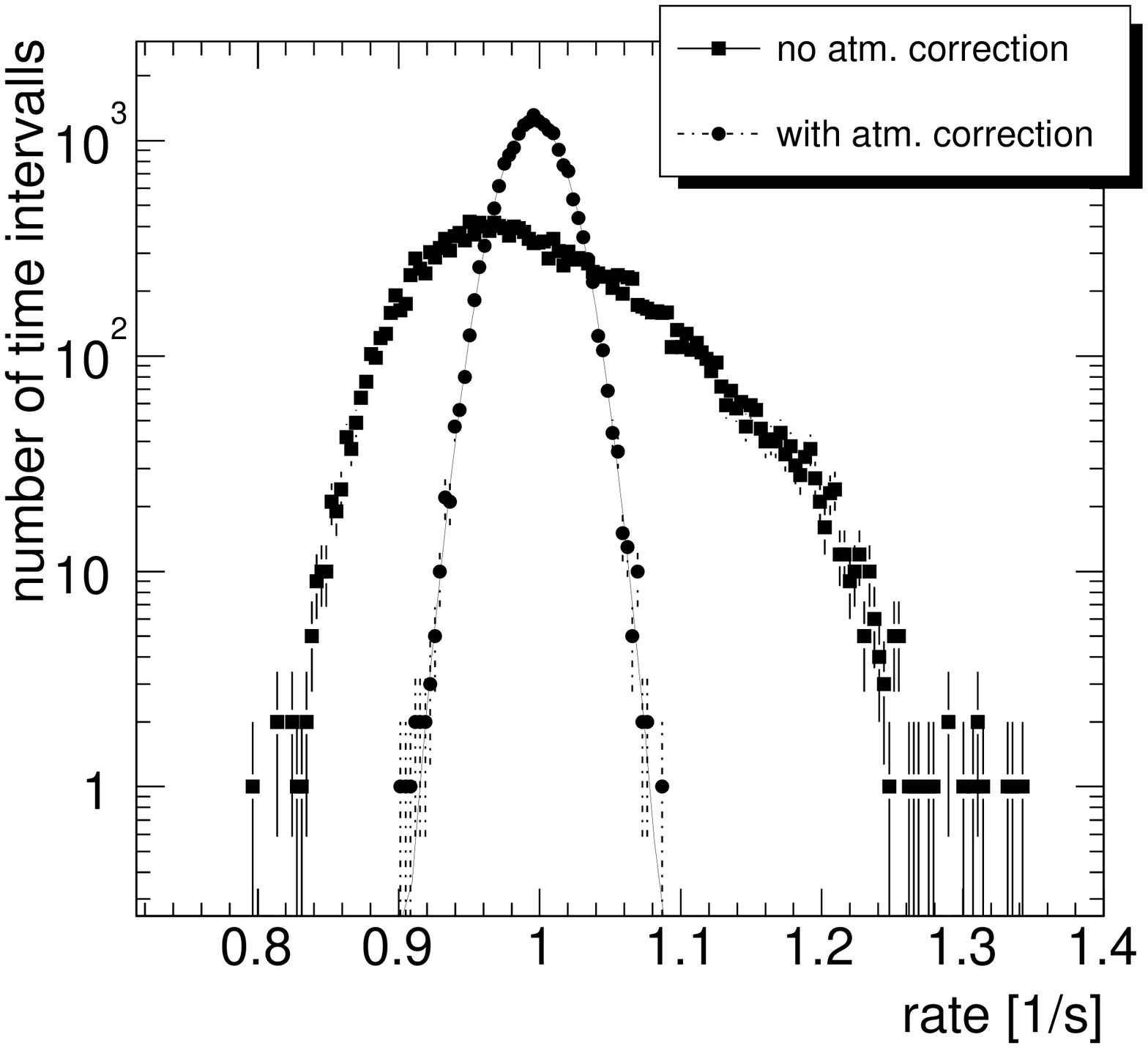}{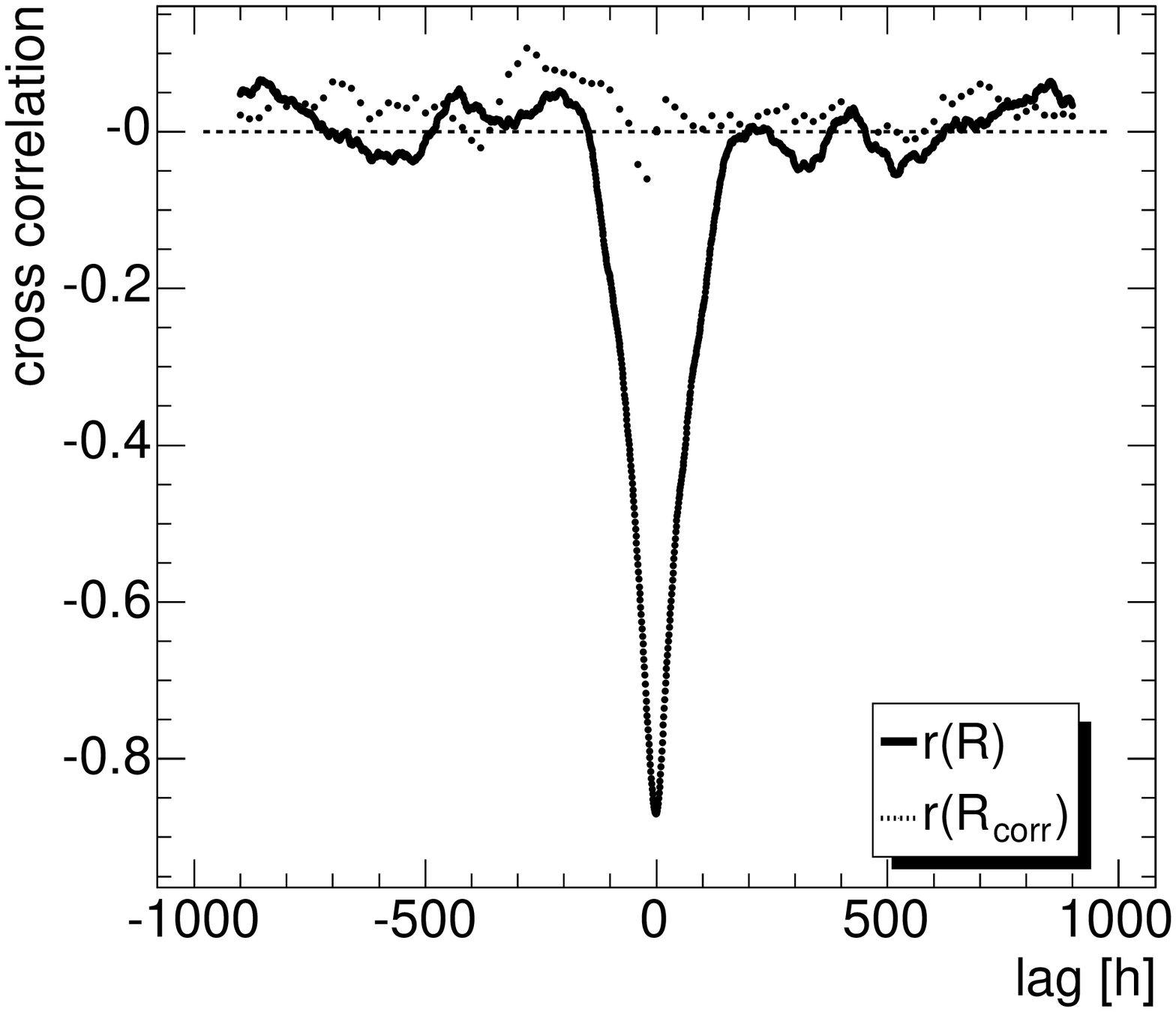}
\caption{\label{fig:atmos}
Left: Distribution of rates with and without correction for atmospheric ground
pressure and temperature.
A fit by a Gaussian function is shown by the line.
Right: Cross correlation between hourly shower rate and atmospheric ground pressure
with ($r(R_{corr})$) and without atmospheric correction ($r(R)$).}
\end{figure}

The influence of the atmospheric ground pressure $P$ and temperature $T$
on the rate of extensive air showers at ground is taken into account 
by a second order polynomial
with additional time dependent corrections $\Delta R_i$ for $i$ 
different configurations of the detector system (e.g.~with slightly different
high-voltages of the detectors):
\begin{eqnarray}
\label{equ:baroKorr}
R(P,T,i)  =  R_0 &+& \Delta R_i \nonumber \\
&+&p_1(P-P_0)+p_2(P-P_0)^2 \nonumber \\
&+&p_3(T-T_0)+p_4(T-T_0)^2 \nonumber \\
&+&p_5(P-P_0)(T-T_0)
\end{eqnarray}
The detector parameters are constant during
each of the $i$ time intervals. 
$R_0 = 0.966$ s$^{-1}$, $P_0 = 1002.95$ hPa , $T_0 = 9.7$\Grad C \
are the long-time mean values of rate, ground pressure,
and temperature.
All parameters $p_{1,2,3,4,5}$ and $\Delta R_i$ are estimated by a fit to
the time dependent rates for the whole interval of four years
and result in:
$p_1 = (-7.21\pm 0.0016)\cdot 10^{-3}$, 
$p_2 = (-3.00\pm 0.13)\cdot 10^{-5}$,
$p_3 = (-3.64\pm 0.0023)\cdot 10^{-3}$,
$p_4 = (-3.19\pm 0.22)\cdot 10^{-5}$, and
$p_5 = (-3.76\pm 0.253)\cdot 10^{-5}$ in units of hPa, \Grad C, and seconds.
The values for the $\Delta R_i$ are between $-2 \times 10^{-2}$
and $2 \times 10^{-2}$ s$^{-1}$.
The correction itself is done for time intervals of 1795 s by subtracting
or adding the necessary number of events calculated
by Equation~\ref{equ:baroKorr}.
Events are chosen randomly from the half-hour intervals to lower the
number of showers.
The events, which have to be added for this correction are
chosen randomly from the set of showers
of the same sidereal day. 
The quality of the correction can be estimated from Figure~\ref{fig:atmos}.
The left figure shows the event rate distributions before and after the
corrections.
The uncorrected rates
reflect the asymmetric distribution of the
atmospheric ground pressure.
The distribution of corrected rates
is compatible with a Gaussian distribution, which is expected
from remaining statistical fluctuations of the event rate.
The right figure shows the cross correlation between rate and ground pressure.
The very strong correlation $r(R)$ visible for the uncorrected
rates vanishes after the correction ($r(R_{corr})$)
with Equation~\ref{equ:baroKorr}.


\section{Results}

\begin{figure}
\plotone{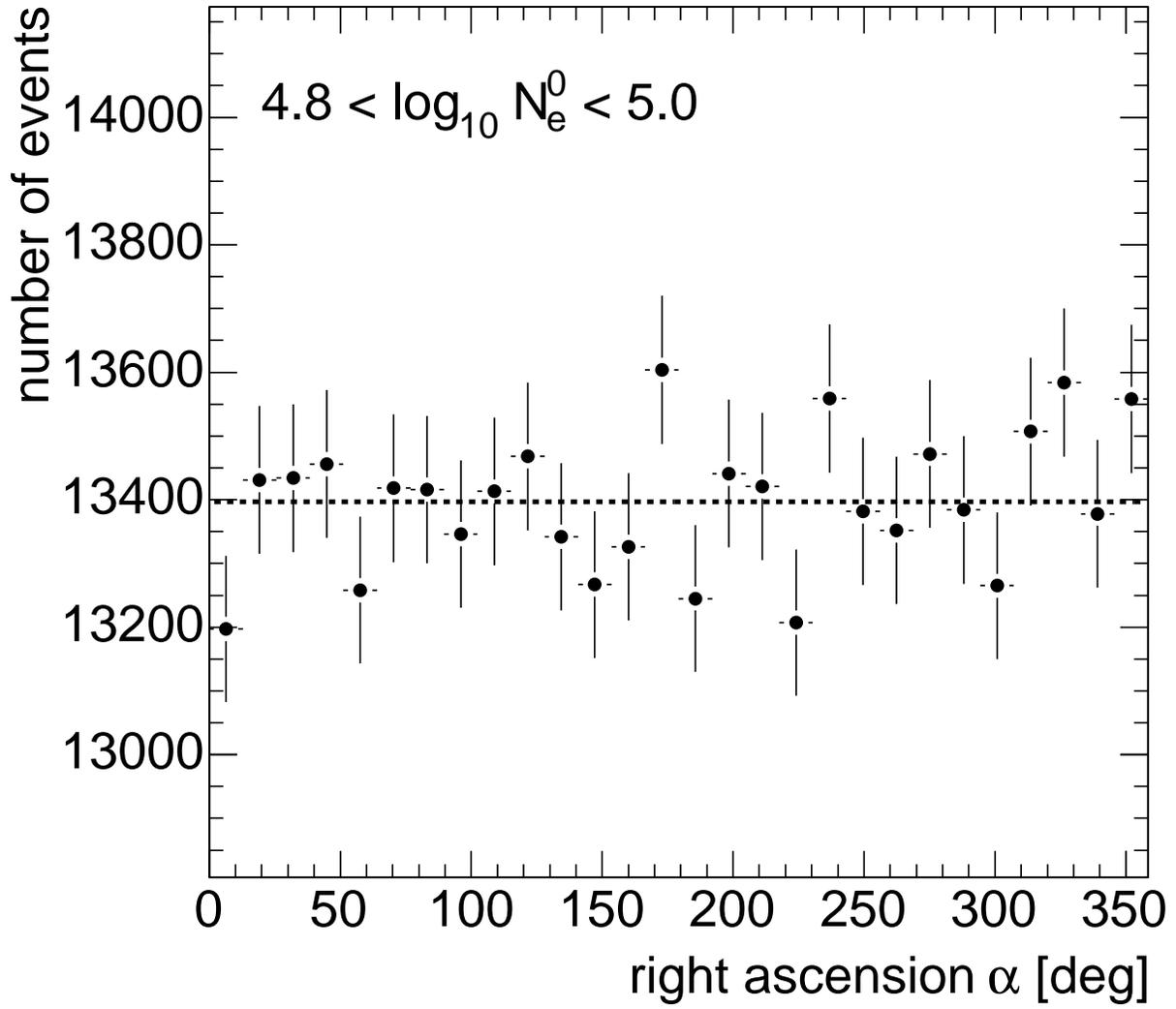}
\caption{\label{fig:RAdis}
Example of a right ascension distribution
after correction for atmospheric ground pressure and temperature
in the stated electron number interval.
The dashed line reflects the mean number of events.
}
\end{figure}

An example for the right ascension distributions of
showers after atmospheric correction
for electron numbers in the interval $4.8 < \log_{10} N_e^0 < 5$ 
is shown in Figure~\ref{fig:RAdis}.
The Rayleigh amplitudes $A$ of the right ascension
distributions for electron numbers
in the range from $\log_{10} N_e^0 = 4$ to $\log_{10} N_e^0 = 6.6$ are
calculated according to Equation~\ref{equ:Ray}.
The lower electron number limit is the efficiency threshold 
of the KASCADE detector field.
As mentioned in section \ref{sec:cuts}, full efficiency is required
in order to minimize  effects of the threshold to the 
amplitudes.
The upper electron number limit at $\log_{10} N_e^0 = 6.6$ is determined by
the small number of events ($\approx 1000$) in this electron number interval.

The showers are sorted in intervals of electron number $\log_{10} N_e^0$
and not in intervals of truncated muon number $\log_{10} N_{\mu,tr}^0$, although the latter one is a 
better estimator for primary energy and less 
dependent on the mass of the primary particle.
The trigger threshold of the detector field is mainly determined by the
number of electrons.
The usage of $N_{\mu,tr}^0$ combined with the requirement of full detection efficency 
would shift the lower
energy threshold of this analysis to energies above $4 \times 10^{15}$ eV.

\begin{figure}
\plottwo{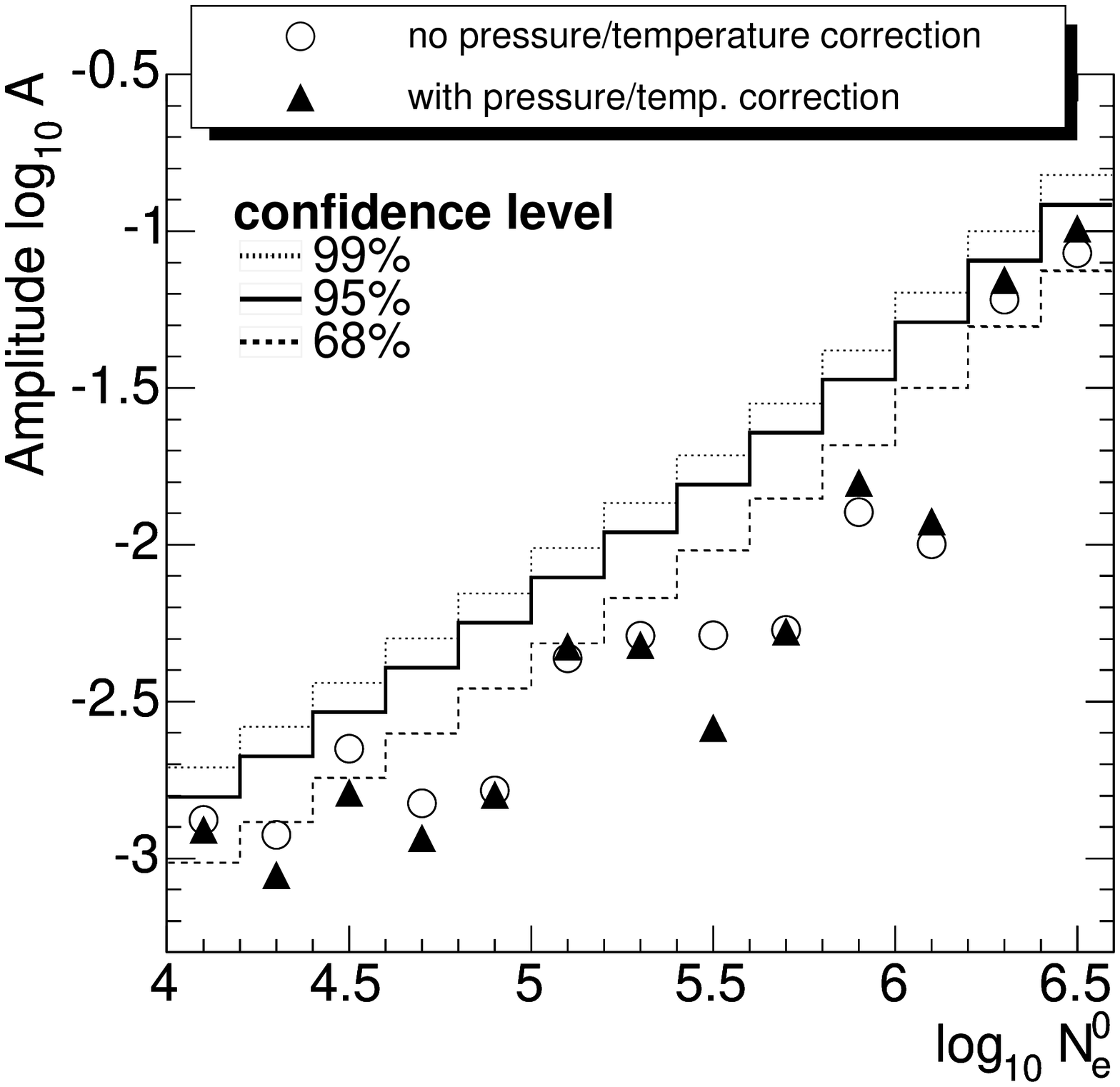}{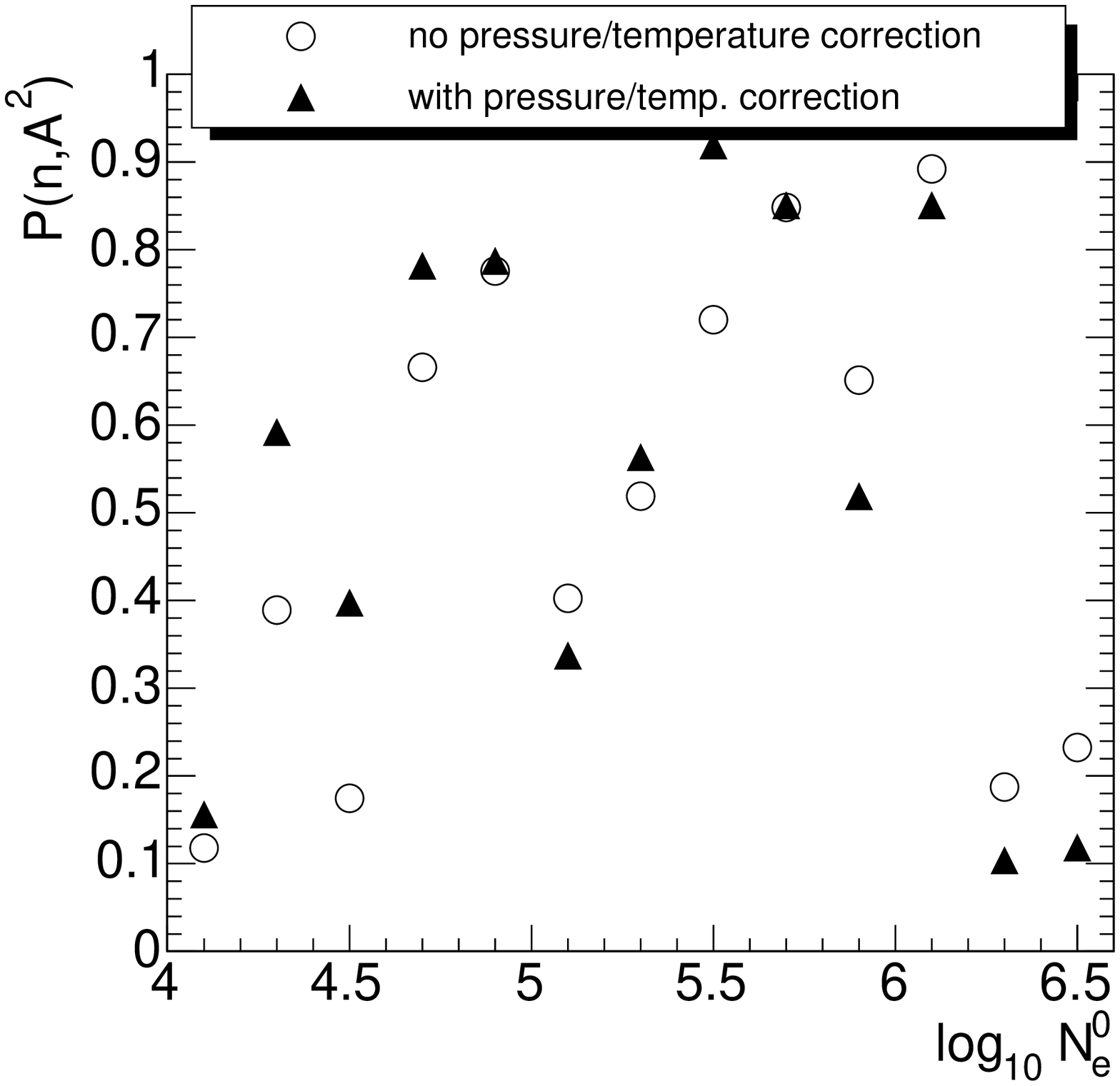}
\caption{\label{fig:RayIntP200}
Result from the Rayleigh analysis for showers without and with
correction for atmospheric ground pressure and
temperature.
Left-hand side: Rayleigh amplitude $A$ vs.~electron number $ \log_{10} N_e^0$.
Right-hand side: Probability $P(n,A^2)$ that the amplitudes plotted
in the left figure are fluctuations from underlying uniform distributions.}
\end{figure}

Figure~\ref{fig:RayIntP200} (left-hand side) 
shows the resulting Rayleigh amplitudes.
The lines indicate the confidence levels for Rayleigh amplitudes 
with probabilities $1-P(>A)$ of 68/95/99 \% respectively.
Assuming a power law with spectral index $\gamma$
for the form of the electron size spectrum, 
the confidence levels are as well power laws with
spectral indices of $-\gamma/2$ (see Equation~\ref{equ:RayP}).
The confidence levels are only a function of the number of events
used in the analysis.
The increase of the confidence levels with electron number
reflects therefore by no means an increase of anisotropy.
Amplitudes which are below the lines indicating the confidence 
levels can be treated as
fluctuations and are of no physical meaning.
All calculated amplitudes are well below the
95\% line.
The fluctuation probability $P(>A)$
of each Rayleigh amplitude 
is shown in Figure~\ref{fig:RayIntP200} (right-hand side).
The probabilities are all above 5\% .
There are no hints for nonzero Rayleigh amplitudes within the
statistical limits.

\begin{figure}
\plottwo{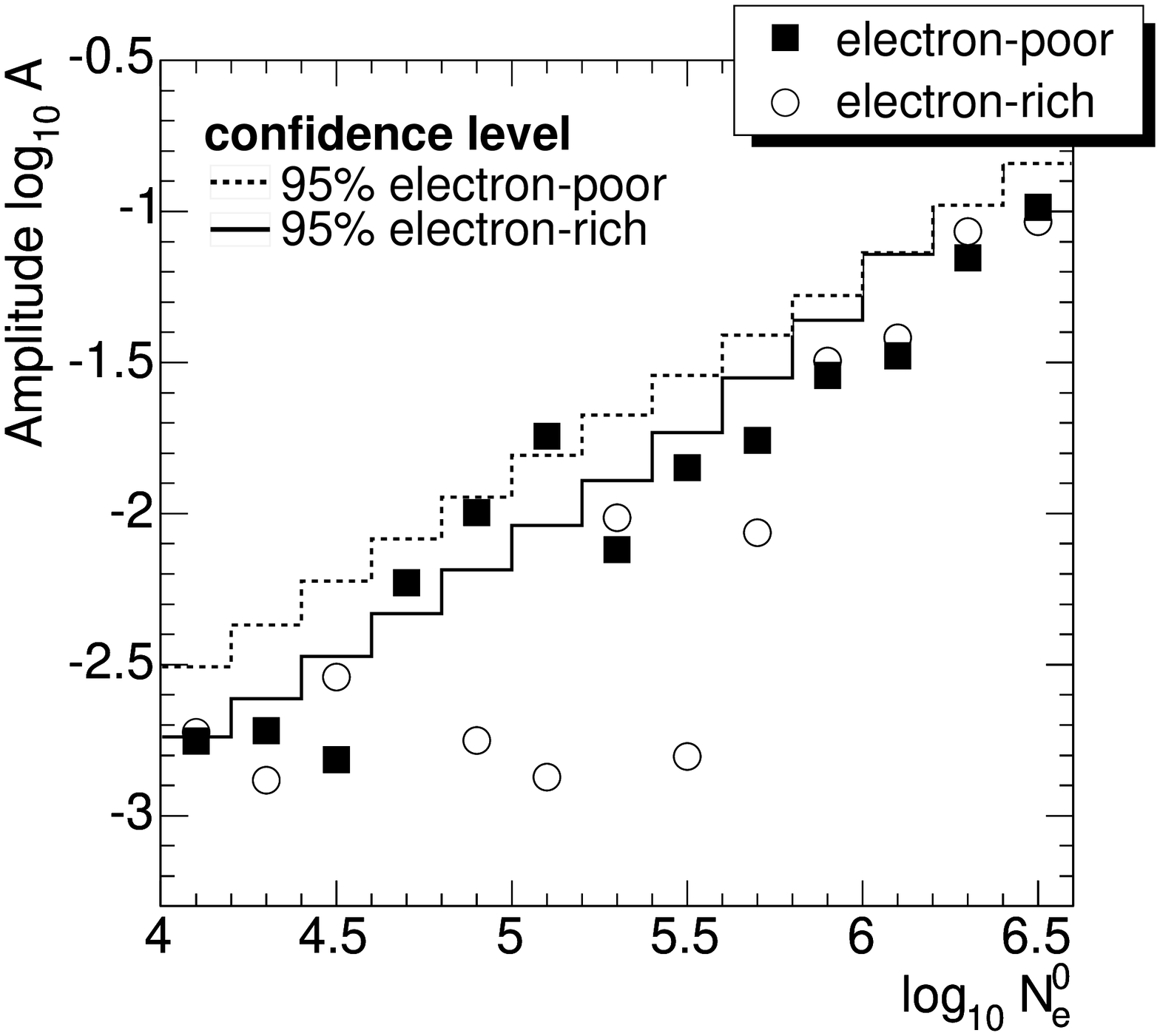}{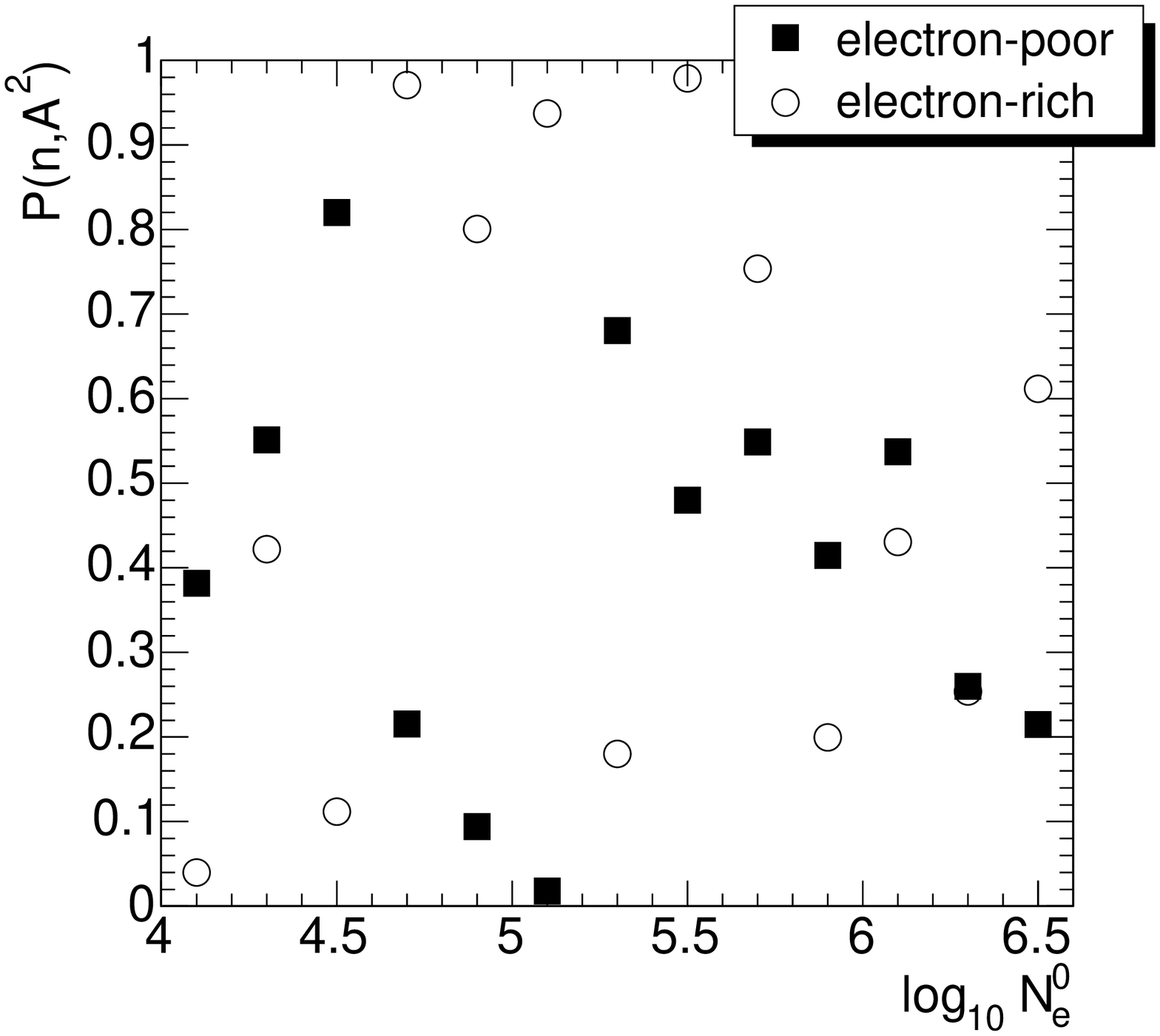}
\caption{\label{fig:RayIntPFE}
Result from the Rayleigh analysis for data sets of predominately
light and heavy primaries.
Left-hand side: Rayleigh amplitude $A$ vs.~electron number $\log_{10} N_e^0$.
Right-hand side: Probability $P(n,A^2)$ that the amplitudes in the left figure
are fluctuations from underlying uniform distributions.}
\end{figure}

The results of the two subsets of data containing electron-rich
and electron-poor showers are shown in Figure~\ref{fig:RayIntPFE}.
No correction for ground pressure $P$ and temperature $T$ is applied in this
case.
The variations of $P$ and $T$ alter beside the detection rate also the number
of electrons and muons in a shower and therefore the effect of the separation line
(Equation~\ref{equ:gammasSepCut})
between electron-poor and electron-rich showers.
A further correction would require detailed information about 
the influence of atmospheric variations on $N_e$ and $N_{\mu,tr}$,
which is beyond the scope of this analysis.
Only a detection of significant amplitudes would require such 
further steps.
As can be seen from Figure~\ref{fig:RayIntPFE}, no anisotropy
can be deduced from the calculated amplitudes
and fluctuation probabilities.
The most prominent amplitude at
$5<\log_{10} N_e^0 < 5.2$ has a significance of $\sigma=2.2$.
The intersection of the confidence levels of electron poor and rich
showers is due to the increasing fraction of heavy primary cosmic rays
with increasing primary energy in the region of the knee
\citep{Antoni:2002}.

Additionally to the results presented in Figures \ref{fig:RayIntP200} and
\ref{fig:RayIntPFE}, an analysis of the data set with different definitions
of the electron number intervals and for showers above the knee sorted
by truncated muon numbers yielded the same result of no significant amplitudes.

\begin{figure}
\plotone{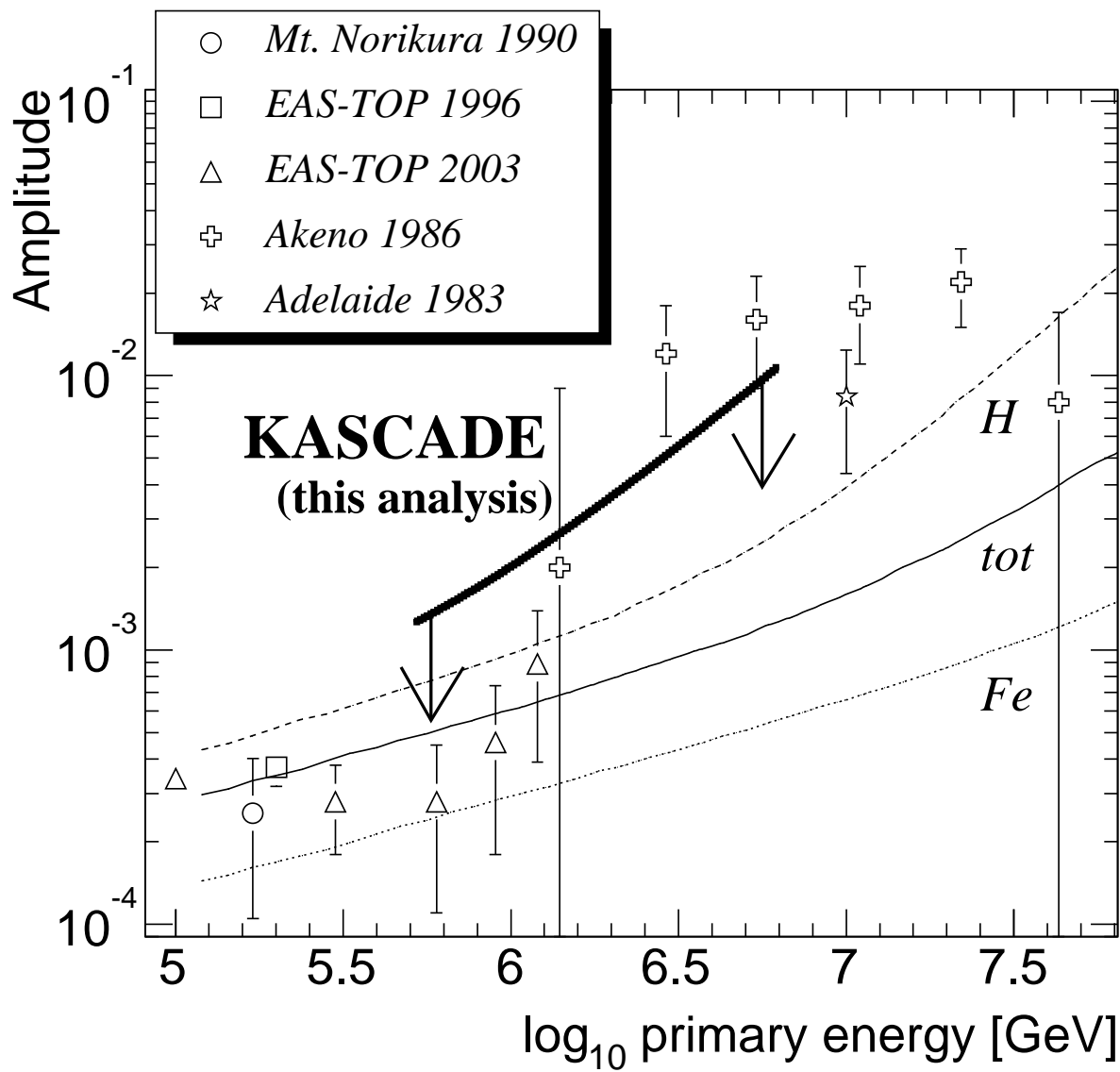}
\caption{\label{fig:results}
KASCADE upper limits (95\%) of Rayleigh amplitudes $A$ vs. primary energy
(bold line)
compared to reported results from literature 
\citep{Nagashima:1990,EASTOP96,Aglietta:2003,Kifune:1986,Gerhard:1983}.
Model predictions from \cite{Candia:2003}
for the total anisotropy and for anisotropies of the proton
and iron component are shown as well (thin lines).
}
\end{figure}

Figure~\ref{fig:results} shows the upper limits on the large scale
anisotropy derived in this analysis in context with results from
other experiments and predictions from the model of \cite{Candia:2003}.
The primary energies of the extensive air showers measured by KASCADE
are determined by a linear transformation
of the particle numbers $\log_{10} N_e^0$ and $\log_{10} N_{\mu,tr}^0$. 
The transformation matrix is determined from CORSIKA simulations
using the hadronic interaction models QGSJET and GHEISHA.
The uncertainty of this simplified primary energy determination
is about 20\%.
The figure shows that the KASCADE upper limits are in the range
of the reported results from other experiments.
The EAS-TOP experiment reported somewhat lower limits
in the energy range below $2\times 10^{15} $ eV. 
The relatively large amplitudes published by the Akeno experiment
are difficult to reconcile with the results of this analysis.
The model calculations, which dependent of course on
several parameters like the source distribution or the strength and
structure of the galactic magnetic field, yield 
amplitudes in the range of $3\times 10^{-4}$ to $2\times 10^{-3}$ in the
energy range of KASCADE.
This is about a factor of 4-10 lower than the upper limits
derived in this analysis.
The contribution from anisotropy measurements towards a solution
of the enigma of the knee are therefore still small.
A significant observation of the anisotropy of separate cosmic ray 
components around and above the knee requires much larger data sets
compared to the presently available.



\acknowledgments

The authors would like to thank the members of the engineering and technical staff of the KASCADE
collaboration who contributed with enthusiasm and commitment to the success of the experiment. 
The KASCADE experiment is supported by the German Federal Ministry of Education and Research
and was embedded in collaborative WTZ projects between Germany and Romania (RUM 97/014) and
Poland (POL 99/005) and Armenia (ARM 98/002). 
The Polish group acknowledges the support by KBN grant no. 5PO3B 13320.

\clearpage

%

%

\clearpage

\end{document}